\documentclass[conf]{new-aiaa}
\usepackage[utf8]{inputenc}
\usepackage{graphicx}
\usepackage{amsmath}
\usepackage[version=4]{mhchem}
\usepackage{siunitx}
\usepackage{longtable,tabularx}
\usepackage{xcolor}
\usepackage{color,soul}
\usepackage{fancyhdr}
\usepackage[nameinlink,capitalize]{cleveref}

\title{Analysis of thermochemical non-equilibrium hypersonic flow over a waverider with uncertainty quantification}

\author{Jeremy Redding\textsuperscript{1,2}, Nick Plewacki\textsuperscript{3}, Himakar Ganti\textsuperscript{1,2}, Luis Bravo\textsuperscript{4}, and Prashant Khare\textsuperscript{1,2}} 
\affil{\textsuperscript{1}Department of Aerospcace Engineering, University of Cincinnati, Cincinnati, OH 45221-0070 USA}
\affil{\textsuperscript{2}Hypersonics Laboratory, Digital Futures, University of Cincinnati, Cincinnati, OH 45206 USA}
\affil{\textsuperscript{3}SURVICE Engineering, Aberdeen Proving Ground, MD, USA}
\affil{\textsuperscript{4}DEVCOM Army Research Laboratory, Aberdeen Proving Ground, MD, USA}

\begin{document}
\maketitle

\begin{abstract}
The objective of this work is to assess the impact of parameter uncertainty on hypersonic aerothermal surface heating predictions in Reynolds-Averaged Navier-Stokes (RANS) simulations using non-intrusive uncertainty quantification (UQ) techniques. RANS-based models are considered indispensable tools in computational fluid dynamics (CFD) analysis for the iterative and cost-effective exploration of innovative design concepts. However, these RANS models heavily rely on empirical constants that often require tuning due to the lack of physical knowledge and complexity of the problem, introducing significant uncertainties that hinder their predictive capabilities. Therefore, this research investigates the influence of the turbulent Prandtl number ($P_{tr}$) uncertainty, that governs the level of shear stress and heat flux present in the  turbulent flow, on key output quantities of interest (QoIs). The US3D hypersonics solver is employed to simulate  aeroheating over a hypersonic waverider configuration using the classical Menter Shear Stress Transport (SST) turbulence model. 
A preliminary study was conducted to demonstrate US3D’s ability to predict the aerodynamic and heat flux parameters with acceptable comparison to experiments for a range of flight conditions. Then, a polynomial chaos expansion (PEC) framework is presented that enables a global sensitivity analysis and forward propagation of uncertainty for a range of $P_{tr}$, generating statistics including skewness and kurtosis of the QoIs. In addition, Sobol indices are calculated to quantify the relative contribution of the $P_{tr}$ to the overall uncertainty in the heat flux and surface pressure outputs. The results provide valuable insights into the underlying aeroheating behavior in RANS simulations under hypersonic non-equilibrium flow conditions over a waverider previously studied at the Arnold Engineering Development Center (AEDC) facility. These findings will inform future design processes and improve the reliability of RANS-based predictions in hypersonic applications.
\end{abstract}

\section{Introduction}

\begin{figure}[hbt!] 
    \centering
    \includegraphics[width=0.75\linewidth]{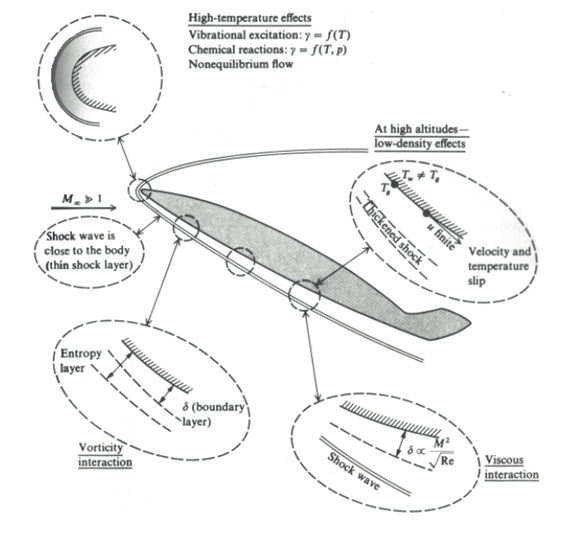}
    \caption{Physical effect characteristics of hypersonic flow around a vehicle \cite{anderson_aiaa_1991}} 
    \label{fig:aerothermal heating}
\end{figure}

A hypersonic aircraft traveling through the atmosphere is subjected to severe aerothermal loads from its harsh surroundings and, in many cases, also from its propulsion systems \cite{KUCHEMANN1965271}\cite{Finley_book_1990}.
Its operating environment is characterized by strong shock waves formed at the leading edges, high-temperature gases, and a viscous boundary layer attached to the whole vehicle surface \cite{candler_arfm_2019}, \cite{cheng1993perspectives},\cite{smith_ft_2021}. Figure \ref{fig:aerothermal heating} 
shows the various fluid dynamic processes around a vehicle in hypersonic flight. Peak heat fluxes generally occur at the stagnation point located in the nose region but also in downstream locations where the boundary layer transitions to turbulence, leading to rapid increases in Nusselt number and associated heating rates \cite{Fu_Karp_Bose_Moin_Urzay_JFM_2021}\cite{benay_aiaa_2006}. Thus, understanding aerothermal loading and the transition to turbulence is of great importance, as it sets the performance requirements for the design of critical components such as thermal protection systems (TPS), control systems, and propulsion \cite{UYANNA_AA_2020}.  \\

Reliable predictions of hypersonic turbulent flow is essential for the design of next-generation high-speed aerospace vehicles \cite{candler_arfm_2019}. One of the main challenges in modeling aerothermal heating is its inherent multiphysics nature involving the presence of strong gas-dynamics effects, such as shocks and compressible boundary layer turbulence \cite{anderson_aiaa_1991}. Strong nonlinear heating generated within the boundary layer leads to a rapid increase in gas temperature, allowing vibrational excitation, chemical dissociation, and ionization dynamics to emerge \cite{Bertin_ARFM_2006}. These thermochemical phenomena can proceed on timescales comparable to those of hydrodynamics, introducing chemical and thermodynamic non-equilibrium. The design of hypersonic vehicles is typically conducted based on computational fluid dynamic (CFD) software tools that employ the Reynolds Averaged Navier Stokes (RANS) approach for turbulence modeling, and in particular closures using linear eddy viscosity \cite{launder_reece_rodi_1975}\cite{menter_aiaaj_1994}\cite{rumsey_jsr_2010}. These modern tools for engineering-level design studies generally rely on fast lower-fidelity models to conduct wide parametric and optimization studies \cite{KOLESNIK2022105622}\cite{BOYCHEV2020105640}. The fidelity of the aerodynamics and aerothermodynamics databases is anchored by using a limited number of RANS solutions to capture the complex physical processes experienced by a vehicle flying at hypersonic speeds. Thus, reliable physics-based models that can simulate the multiphysics processes of aerothermal heating, transition, and other complex gas-surface interactions are urgently needed due to the intractable cost of Direct Numerical Simulation (DNS) or Large Eddy Simulation (LES) approaches that are not yet feasible for design studies generally requiring 100s+ concurrent simulations. 

From an engineering perspective, a reliable modeling approach must be selected to design and optimize any hypersonic aircraft or propulsion systems. One way to improve the accuracy and reliability of existing RANS-based simulation tools is by increasing the understanding of how epistemic model uncertainties (uncertainty due to a lack of knowledge) affect the overall predictions and their sensitivities. Recently, an area that has received renewed attention is the evaluation of the closure coefficients of the turbulence model on aerodynamic predictions \cite{LIU_AA_2018}\cite{ZHAO_IJHMT_2019}. The Reynolds stress model closure coefficients in RANS are selected empirically or heuristically and their sensitivities are not well understood. 
Thus, current turbulence models used in RANS simulation are not guaranteed to perform well for any arbitrary flow and can often fail in flow regimes significantly dissimilar to the ones used in their formulation \cite{ERB2021105027}. Thus, an ability to estimate the epistemic uncertainties behind the model assumptions is essential for their reliable use in design analysis. Therefore, the objective of this work is to develop an approach for the quantification of uncertainty of RANS simulations of aerothermal heating of high-speed aircrafts. To achieve this goal, we consider the problem of thermochemical hypersonic flow over a delta waverider. The RANS simulations were performed using the compressible unstructured finite-volume RANS solver, US3D. The source of uncertainty in the simulation is the turbulent Prandtl number ($P_{tr}$) as that governs the level of shear stress and heat flux present in the turbulent flow and is invoked in the momentum and energy equations. To quantify the influence of uncertainties in the $P_{tr}$ on the prediction of the quantity of interest, perturbations of the values computed by the RANS models are introduced in the simulations. A preliminary study is first conducted to demonstrate US3D’s ability to predict the aerodynamic and heat flux parameters with acceptable comparison to experiments for a range of flight conditions. Then, a polynomial chaos expansion (PEC) framework is presented that enables a global sensitivity analysis and forward propagation of uncertainty for a range of $P_{tr}$, generating statistics including skewness and kurtosis of the QoIs. Further, Sobol indices are calculated to quantify the relative contribution of the $P_{tr}$ to the overall uncertainty in the heat flux and surface pressure outputs.  The remainder of the paper is organized as follows: Section 1 presents the motivation and overall objective of this work, Section 2 presents the computational framework including the coupling of the computational fluid dynamics solver and the Dakota statistical package, Section 3 presents the results and discusses the impact of parameter uncertainty on aerothermal heating, and finally Section 4 presents the summary and future work. 
.


\section{Methods}

\subsection{Reynolds Averaged Navier Stokes (RANS) solution}
In this work, the US3D commercial hypersonic solver was used to analyze the Mach 10 waverider subject to a realistic mission profile. US3D is a massively parallel state-of-the-art computational fluid dynamics research and analysis tool originally developed as a collaboration between the University of Minnesota, NASA Ames, and VirtusAero Inc, and currently distributed and maintained by VirtusAero Inc. US3D is parallelized using message passing interface (MPI) libraries and has been deployed on virtually all Department of Defense (DoD) High Performance Computing platforms showing excellent scalability across 1000’s of computer cores. It solves the conservative form of the unsteady compressible Navier-Stokes equations on a finite-volume unstructured mesh, using a range of high-order low-dissipation fluxes. Turbulence, heat transfer, and chemically reacting flow is handled using a range of Reynolds Averaged Navier Stokes (RANS), and Large Eddy Simulation (LES) models, conjugate heat transfer, and finite-rate chemistry methods. Its accuracy has been extensively demonstrated over the years on a variety of hypersonic configurations \cite{candler2015development}.
Thus, the solver is tailored to model supersonic to hypersonic flows and can handle the interactions that include the evolution of strong shocks, turbulence, kinetics, plasma dynamics, and ablation physics, to name a few. 

\subsection{Uncertainty Quantification}

One of the goals of the present study is to quantify the uncertainty in the predictions of the high-enthalpy hypersonic flow over a waverider. The goal of uncertainty quantification is to determine the distribution of the response of a model (say, Y) to uncertainty in the distribution of an input variable (say, X). 
The distribution of response variable Y can be in the form of data sets that include first- and higher-order moments, as well as probability density functions, cumulative density functions, and simple correlation coefficients.

Polynomial chaos expansion (PCE) is an effective tool that has been widely used to quantify uncertainty in models \citep{mcclarren2018uncertainty} parameterized by both dependent and independent random variables. In CFD simulations, a common implementation is a special form of PCE deemed non-intrusive PCE (NIPCE). This method looks only at the inputs and outputs of the simulation, avoiding modification to the simulation techniques. In NIPCE, randomly sampled variables are input into the simulation(s), and the input-output map is built (defined by the orthogonal polynomial expansion chosen), enabling the approximation of expected values and variance of the output values.


In the most general sense, PCE takes the form:

 \begin{equation}
     \alpha^*(\vec{x}, t, \vec{\zeta}) = \sum_{j=0}^{\infty} \alpha_j(\vec{x},t) \Psi_j(\vec{\zeta})
     \label{eq:pcegen}
 \end{equation}
 
where $\alpha^*$ is the random process of interest (i.e., the simulation output based on random inputs), $\Psi_j$ is an element of orthogonal polynomials, and $\alpha_j$ are the coefficients for the expanded polynomial. 
$x$ and $t$ are the unmodified random inputs, and $\zeta$ is the transformed variable dependent on $x$, the transformation for which depends on the choice of polynomial/distribution type. The quantities contained within $\alpha_j$ provide details of the statistical quantities of interest.  


We focus on the turbulent Prandtl number as a random variable, which conceptually provides insight into the differences between turbulent transport of momentum and energy. A non-unity $P_{tr}$ would indicate differences in the temperature and velocity profiles of a system. High flow velocities encountered in hypersonic flight may cause high temperatures near the surface of a vehicle, leading to high rates of heat transfer into the wall/boundary. Estimation and use of a constant global turbulent Prandtl number in such simulations may cause an overestimation of the heat transfer in a system. \citep{xiao2007role} Literature indicates that local $P_{tr}$ has a significant influence on wall heat flux in hypersonic flows, especially in situations where shock wave boundary layer interactions (SWBLI) are important \citep{roy2018variable}.


Given the uncertainty in the experimental prediction of global $P_{tr}$, and given that it has 
a significant effect on the heat flux near the wall of hypersonic vehicles, it is useful to explore the sensitivity of model predictions to $P_{tr}$. Furthermore, with the present interest in waverider vehicle optimization and utilization for hypersonic flight, the application of this uncertainty study to an AEDC waverider will give valuable insight into sensitivity and uncertainties in measurements on similar vehicles, enabling better alignment between simulation and experiment, and more rapid exploration of flight profiles. 

    To accomplish this goal, we performed simulations to match the wind tunnel operating conditions of the AEDC waverider Run 2917.  To determine the influence of turbulent Prandtl number $P_{tr}$ on the heat flux and other near-wall flow properties, a set of 20 deterministic simulations are run, each with a single value falling in the range of $P_{tr} = [0.7:0.9]$, selected specifically to include the average value of the turbulent Prandtl number, which is estimated around $P_{tr_{avg}} = 0.85$ given the theoretical and analytical calculations demonstrated in \citet{li2019turbulent}, for an increasing Peclet number flow. 

For the PCE analysis, we must choose an appropriate orthogonal polynomial expansion series to proceed with a proper uncertainty analysis. The choice of this polynomial is based on their input distribution. Four of the primary orthogonal polynomials used in this method, along with their defined input distributions, are given in \autoref{tab:polynomials}. 

\begin{table}[h]
    \centering
    \begin{tabular}{c|c|c}
        Input random variable distribution & Orthogonal Polynomial & Range \\
        \hline
        Normal & Hermite & ($-\infty$, $\infty$) \\
        Uniform & Legendre & $[-1,1]$ \\
        Beta & Jacobi & $[a,b]$ \\
        Gamma & Laguerre & $[0,\infty)$ 
    \end{tabular}
    \caption{Orthogonal polynomials and the different input distributions and ranges that correspond to them}
    \label{tab:polynomials}
\end{table}
In our case, since we do not have a precise definition of the distribution of this variable, we begin by assuming that each of the values of $P_{tr}$ given are equally likely to occur. This may be more appropriate than a normal distribution, as the range is close to the average term suggested in the literature. However, more data would be necessary to confirm or decide on whether a different distribution is more appropriate. 

In a uniform distribution of random variables, the random variable is known to be transformed into and from an interval [a,b]. In our case, following the efforts of \citet{mcclarren2018uncertainty}, this range is mapped to [-1, 1]. This transformation from the random variable to the mapped input is as follows:

\begin{equation}
    x = \frac{(b-a)}{2}\zeta + \frac{(a+b)}{2}
\end{equation}

And from the mapped inputs back to the original random input variable:

\begin{equation}
    \zeta = \frac{a+b-2x}{a-b}
    \label{eq:transform_x_to_zeta}
\end{equation}

The calculation and expected values and variance then proceed as follows:

\begin{enumerate}
    \item Using \autoref{eq:transform_x_to_zeta}, transform turbulent Prandtl numbers so that they exist on the Legendre range [-1,1], where the min and maximums defined by $a$ and $b$, respectively, are the min and maximums of the turbulent Prandtl number samples selected. 
    \item Build Legendre polynomials, using zeta, as in \autoref{tab:legendrepolynomials} 

    \begin{table}[h]
    \centering
    \begin{tabular}{c|c}
        Order & Legendre polynomial \\
        \hline
        L0 & 1 \\
        L1 & $\zeta$ \\
        L2 & $3\zeta-1$ \\
        L3 & $5\zeta^3 - 3\zeta$ \\
        L4 & $35\zeta^4 - 30\zeta^2 + 3$ \\ 
        L5 & $63\zeta^5 - 70\zeta^3+ 15\zeta$
    \end{tabular}
    \caption{Legendre polynomials}
    \label{tab:legendrepolynomials}
\end{table}

    \item Construct the matrix of polynomials $\psi$ by \begin{equation}
        \Psi = \{L_0, L_1, L_2 ... L_p\}    \end{equation}, so long as the L values based on each zeta are column vectors
    \item Read in one row of data at a time, corresponding to a single measured point of pressure or heat flux, and set equal to $\alpha^*$ from \autoref{eq:pcegen}
    \item Solve the system in \autoref{eq:pcegen}, the solution to which is provided by the linear system 
    \begin{equation}
        \alpha_j = [\Psi^T \Psi]^{-1} \Psi^T \alpha^*
    \end{equation}
    \item Now, we have the coefficients of the PCE, which contain a significant amount of information about the system. The mean of the data at that point is given in $\alpha_j(0)$, or the first indexed value in the coefficients $\alpha_j$
    \item The standard deviation of the datasets are found by the following equation, where we only consider the values subsequent to the first indexed value of the coefficients array ($\alpha_j$): 
    \begin{equation}
        \sigma^2 = \sqrt{\sum{\alpha_j^2*LSN}}
    \end{equation}
    And the value $LSN$ is defined as the squared norm of the Legendre polynomials with weight:
    \begin{equation}
        dw(\zeta) = \frac{d\zeta}{2}
    \end{equation}

    and are given as in \autoref{tab:legendrepolynomialweights}:

\begin{table}[h]
    \centering
    \begin{tabular}{c|c}
        Order & Squared norm values \\
        \hline
        0 & 1 \\
        1 & 1/3 \\
        2 & 4/5 \\
        3 & 4/7 \\
        4 & 64/9 \\ 
        5 & 64/11
    \end{tabular}
    \caption{Squared norms for Legendre Polynomial}
    \label{tab:legendrepolynomialweights}
\end{table}
    
\end{enumerate}

We first investigate a more straightforward statistical analysis of the simulation and then compare those basic statistical values to the employment of the PCE method as a basic proof that our PCE framework is working well towards applications using deterministic data.  


\section{Results and Discussions}

\subsection{Benchmark analysis and comparison to experiments}
\begin{figure}[hbt!] 
    \centering
    \includegraphics[width=0.65\linewidth]{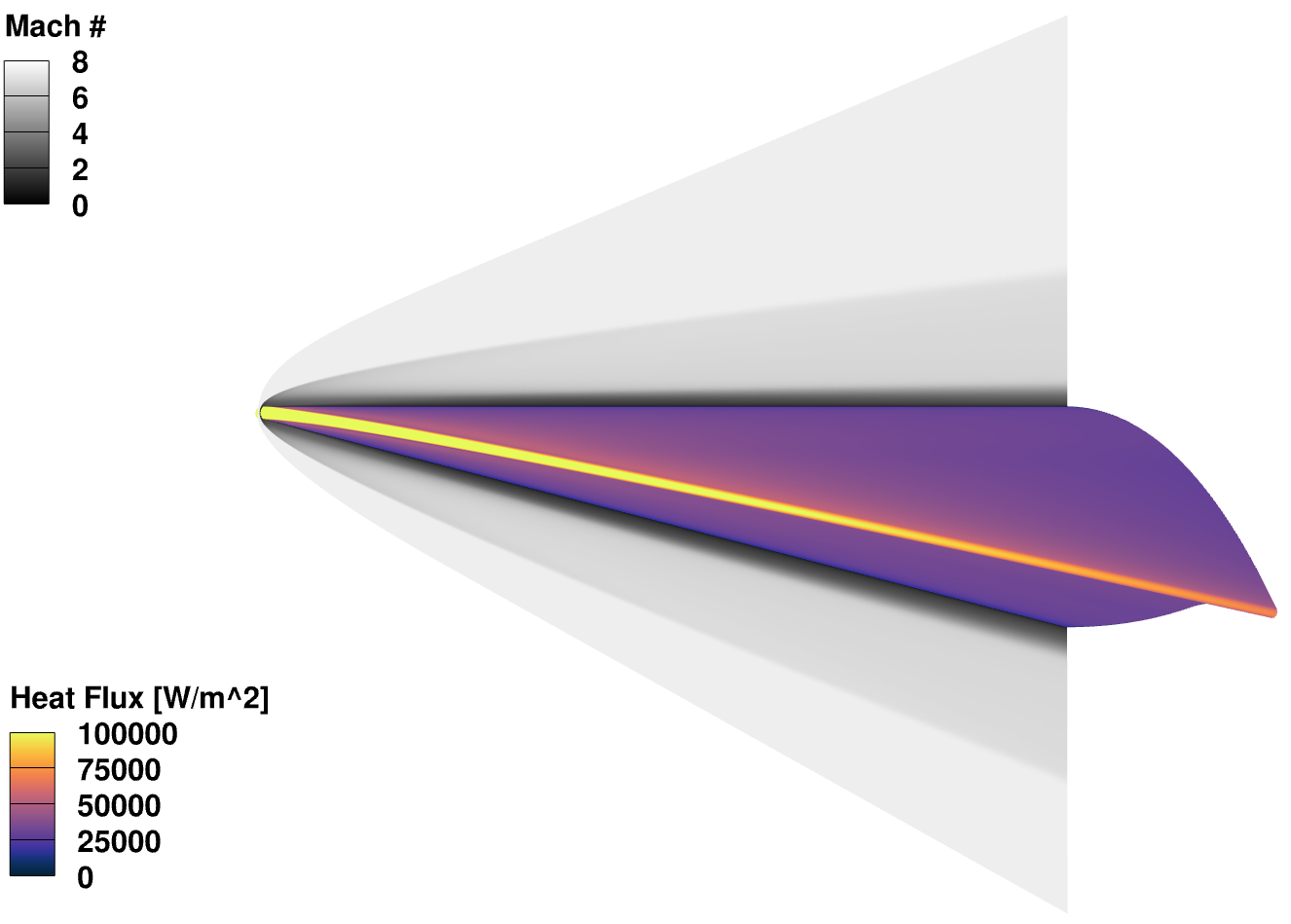}
    \caption{Visualization of US3D aerothermal fields over AEDC hypersonic waverider at AoA = -5 deg and $Pr_{tr}$=0.79} 
    \label{fig:US3Dflowresults}
\end{figure}

To establish the computational method as a reliable data source, simulations were performed to compare with an experimental test campaign carried out in the AEDC Wind Tunnel Number 9 involving a Mach 14 waverider commonly known as the AEDC Waverider in a stinger configuration \cite{norris2006mach}. This waverider is 1 m long with a maximum span of 0.41 m. One portion of this campaign submitted the Mach 14 waverider to nominally Mach 8 conditions and varied the angle of attack. Pressure probes and thermocouples on the surface of the test article allow analysis of the waverider performance under varying conditions and also facilitate comparisons with simulated data. More details on the experimental conditions can be found in \citet{norris2006mach}. US3D simulations using this geometry were performed with a 5-species kinetic model and assumed that the walls were at an isothermal temperature of 300K. The Menter-SST model was used for turbulence closure. Figure \ref{fig:US3Dflowresults} shows a visualization of the aerothermal flow fields and the surface heat flux at AoA = -5 deg. No wake regions or mounting bodies were modeled in the simulations, so the domain was cut off at the rear end of the geometry. The LINK3D meshing tool, developed by GoHypersonics LLC, was used to develop a high-quality structured-style all-hex grid. The simulations were computed using an implicit DPLR time integration method, which followed a prescribed CFL ramping list starting at 0.005 and ending at 1000. Data were taken after the density residual had decreased 10 orders of magnitude from initialization. 

\begin{figure}[hbt!]
    \centering
    \includegraphics[width=0.48\textwidth]{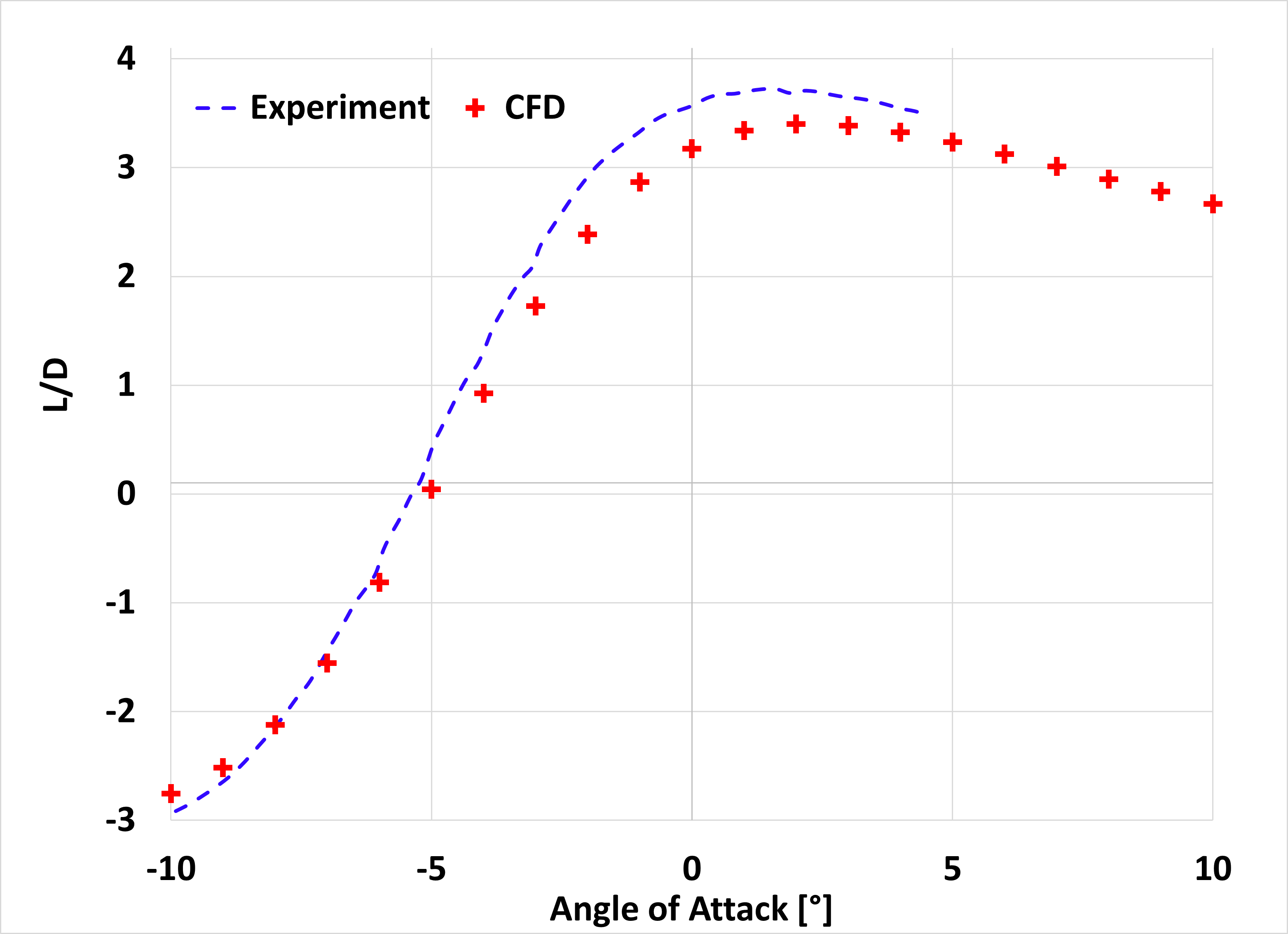}
    \includegraphics[width=0.48\textwidth]{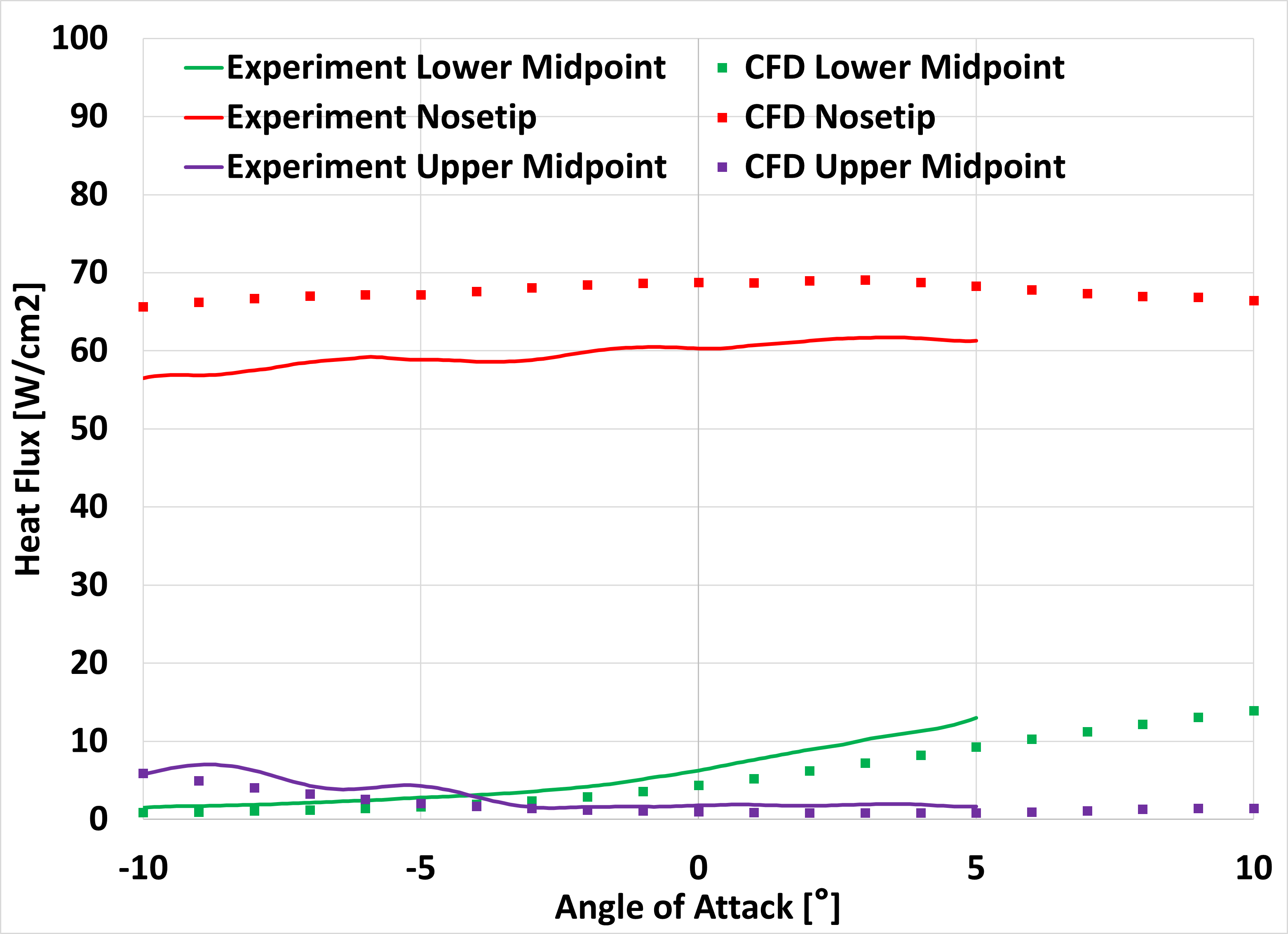}
    \caption{Comparison of Experimental data from AEDC Run 2917 and turbulent CFD simulations showing (left) Lift to Drag ratio and (right) surface heat flux.}
    \label{fig:benchmark1}
\end{figure}

Figure \ref{fig:benchmark1} shows the comparison between our simulations and the experimental data for Run 2917. The lift-to-drag ratios (shown on the left) show excellent agreement between the measured and calculated values at the entire range of angles of attack considered in this campaign. The image on the right shows heat flux values on the surface of the waverider versus angle of attack for three locations: the midpoint of the lower surface centerline, the nosetip, and the midpoint of the upper surface centerline. The nose shows that CFD overestimates the heat flux, which may be due to uncertainties in the nose geometry, or due to the instrumentation uncertainties (exact location of the thermocouple, etc.). Further downstream, the heat flux on the upper and lower surfaces shows excellent agreement between the model and the test data for all angles of attack. 

In addition to the figures shown above, a grid sensitivity study was conducted to ensure the adequacy of the grid to resolve flow phenomena of interest. Two grids consisting of 139.8 million and 50.7 million cells were used to identify the sensitivity of the L/D and heat flux. The 0\textsuperscript{o} angle of attack case showed 
less than 0.2 and 0.6 \% differences in the lift-to-drag ratio and nosetip heat flux, respectively. The centerline heat flux values showed a difference of less than 0. 1\% across the length of the waverider. As a result, the rest of the calculations were performed with the 50.7 M grid (including the results shown in \autoref{fig:benchmark1}).

\subsection{Effect of turbulent Prandtl number}
\begin{figure}[hbt!]
    \centering
    \includegraphics[width=0.48\textwidth]{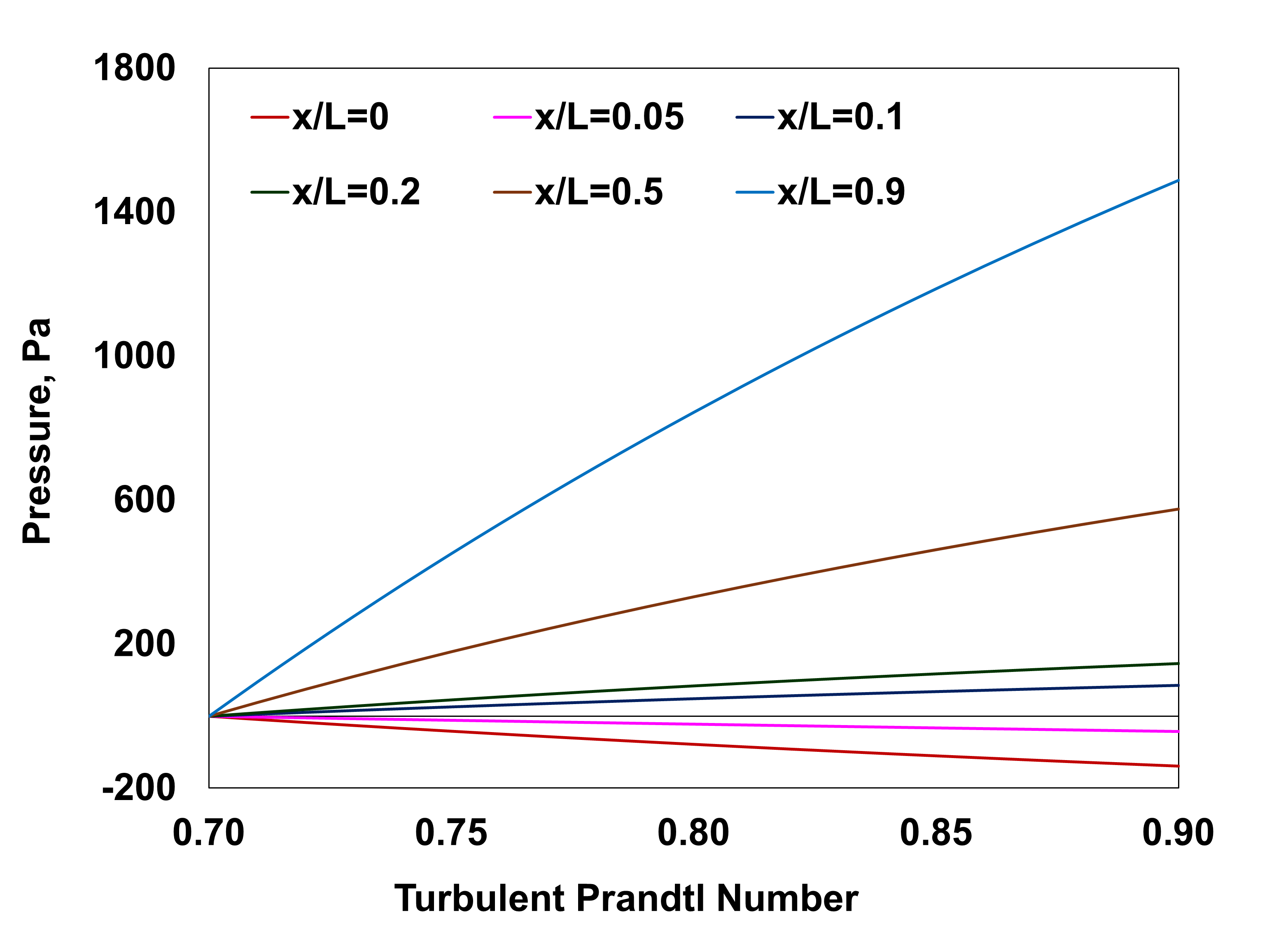}
    \includegraphics[width=0.48\textwidth]{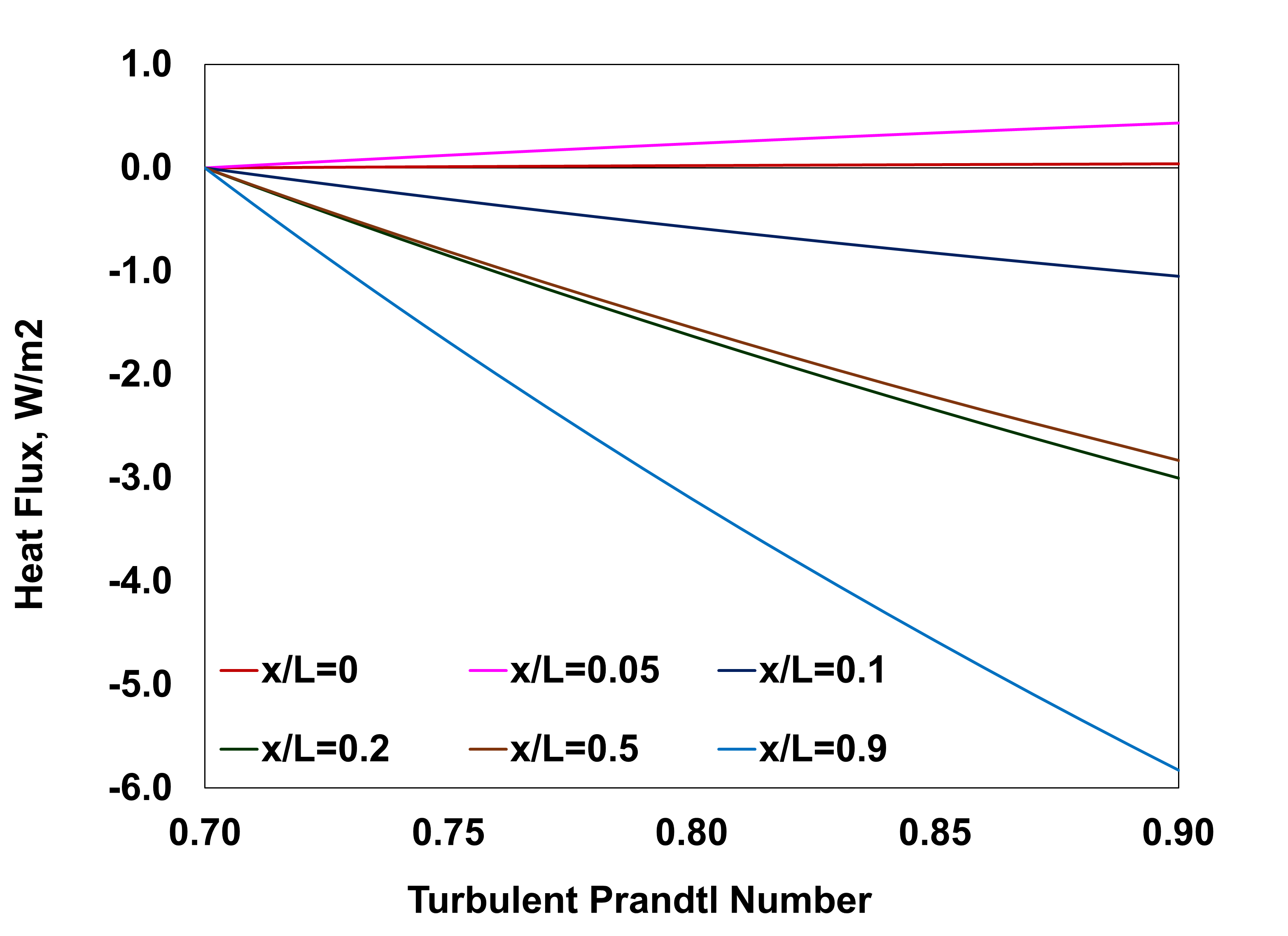}
    \caption{Pressure (left) and heat-flux (right) variations with turbulent Prandtl number at mid-plane locations of AEDC waverider.}
    \label{fig:midplanetrends}
\end{figure}

A parametric study was carried out by varying the turbulent Prandtl number from $0.7$ to $0.9$ in increments of 0.01 in the present study. Figure \ref{fig:midplanetrends} shows the variation of pressure and heat flux for the range of turbulent Prandtl numbers of interest at various midplane locations. Increasing the turbulent Prandlt number results in an increase in pressure difference compared to the pressure value for Prandlt number of $0.70$, at the nosetip and $x/L = 0.05$, while the opposite is observed at the downstream locations of $x/L = 0.50$ and $x/L = 0.90$. The increase in pressure corresponds to a decrease in heat flux at the nosetip and $x = 0.05$ locations, and it is higher at the other downstream locations. The pressures see an significant decrease along the mid-plane as the Prandtl number is increased, while the heat-flux sees a reversed trend and decreases as the Prandtl number is increased. 

\subsection{Sensitivity analysis and higher-order statistics}
Sensitivity is defined as the ratio of the standard deviation of pressure or heat-flux to the standard deviation turbulent Prandtl number, for the sample set described in the previous section, and they are given as
\begin{equation}
\left( {\frac{{\sigma _p }}{{\sigma _{\Pr } }}} \right), \left( {\frac{{\sigma _q }}{{\sigma _{\Pr } }}} \right)
\end{equation}
These coefficients are used to identify the locations with significant changes in pressure and heat-flux values and quantify them for further analysis or design corrections. Sobol indices are similarly defined based on their definition to capture the effects of the independent variable over the entire result as
\begin{equation}
\left( {\frac{{\sigma ^2 _{\Pr } }}{{\sigma ^2 _p }}} \right), \left( {\frac{{\sigma ^2 _{\Pr } }}{{\sigma ^2 _q }}} \right)
\end{equation}
\begin{figure}[hbt!]
    \centering
    \includegraphics[width=0.48\textwidth]{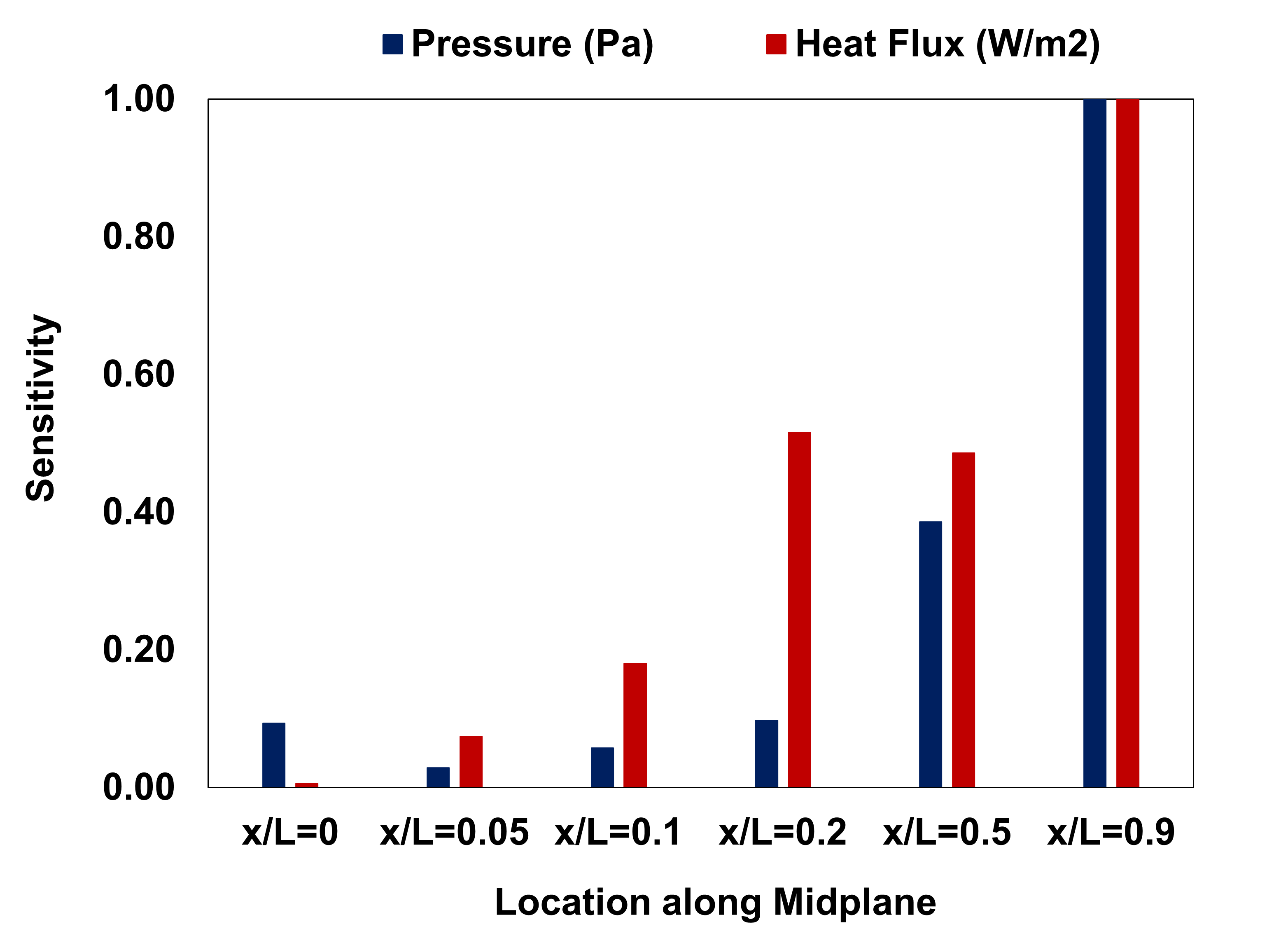}
    \includegraphics[width=0.48\textwidth]{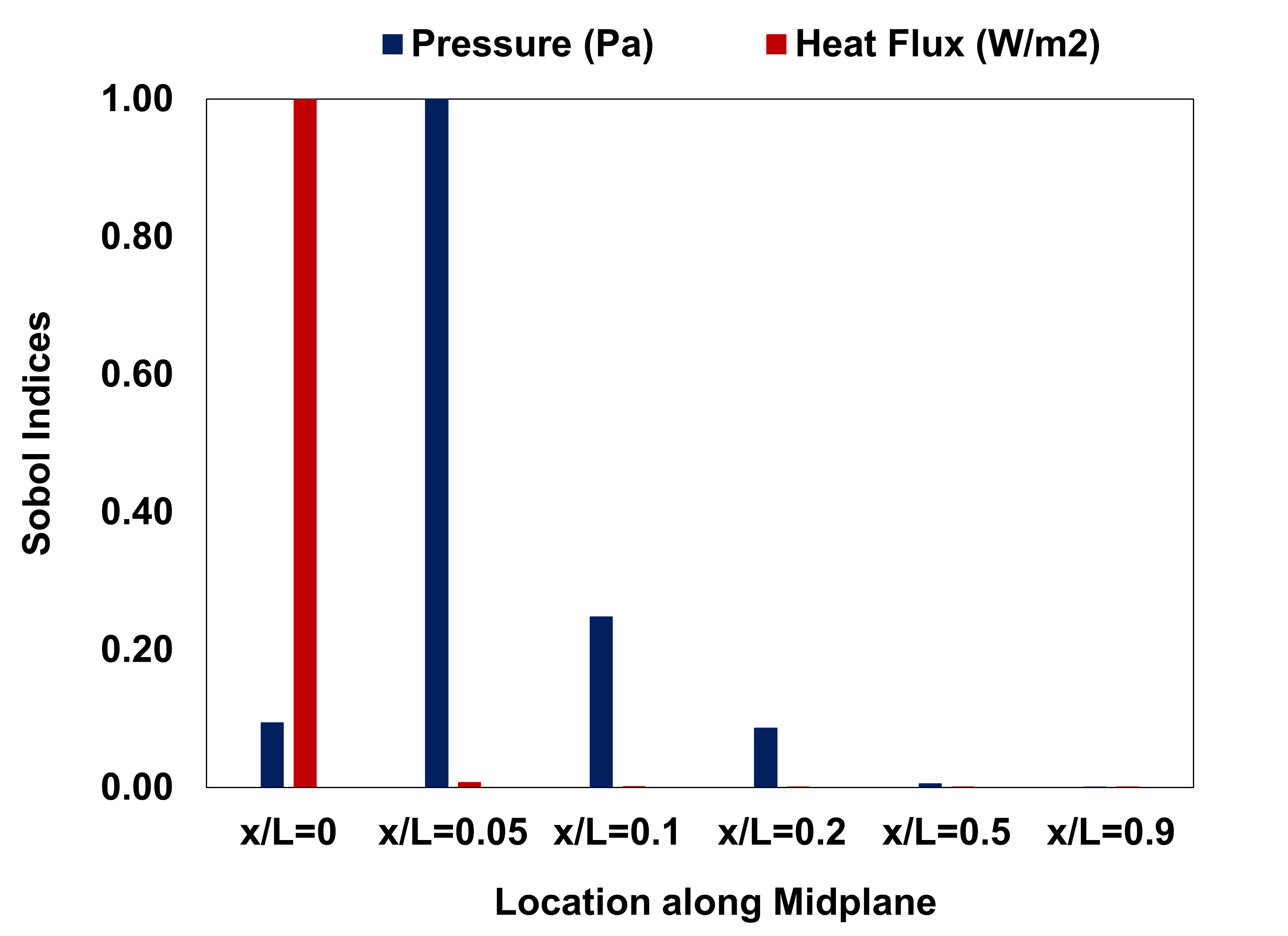}
    \caption{Sensitivity (left) and Sobol Indices (right) for pressure and heat-flux at mid-plane locations of AEDC waverider.}
    \label{fig:midplanestats}
\end{figure}
The sensitivity of pressure and heat flux at midplane locations to the variation of turbulent Prandtl number is shown in figure \ref{fig:midplanestats}. The pressure at the immediate downstream location of $x/L = 0.05$ from the nosetip is the most sensitive to the turbulent Prandtl number identified in the sensitivity plot, and this is further reinforced by the Sobol indices for the pressure at that location that is close to $1$ for that location. The heat flux at that location is most sensitive to the turbulent Prandtl number at the $x/L = 0.2$ location, which is downstream along the mid plane. The Sobol index for that location reflects this - however in comparison to pressure, the heat-flux is not as sensitive to Prandtl number.
\begin{figure}[hbt!]
    \centering
    \includegraphics[width=0.48\textwidth]{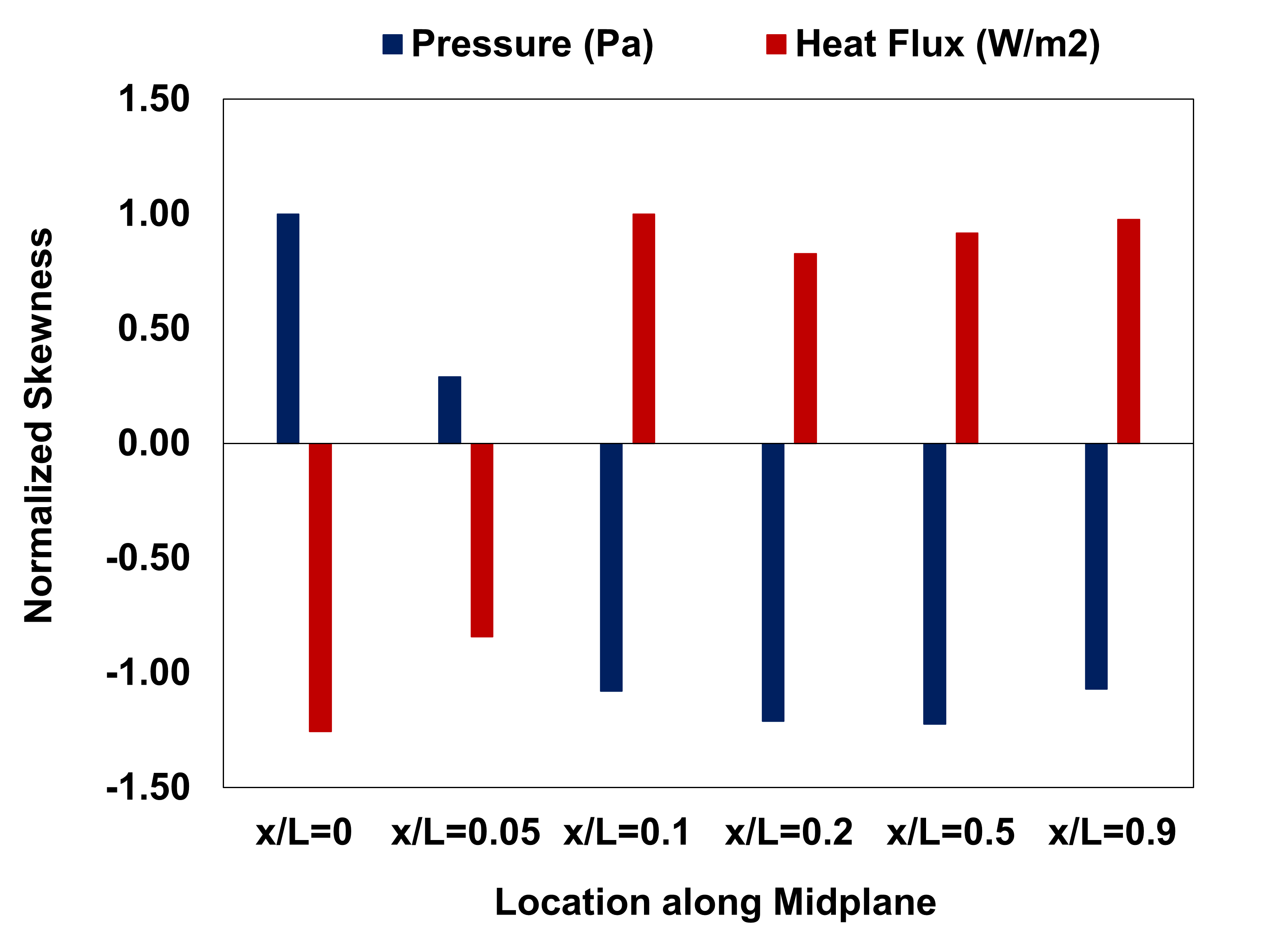}
    \includegraphics[width=0.48\textwidth]{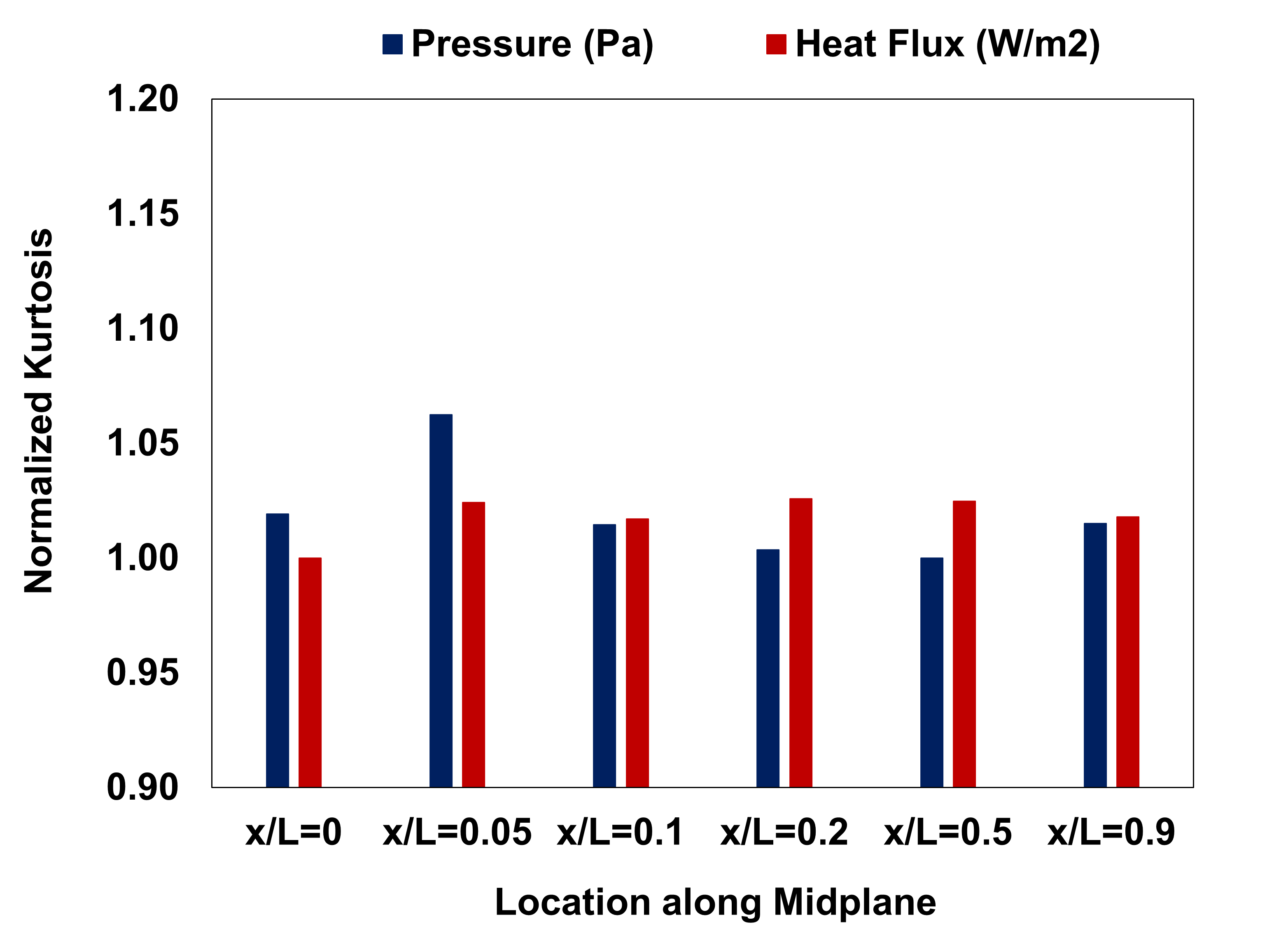}
    \caption{Skewness (left) and Kurtosis (right) for pressure and heat-flux at mid-plane locations of AEDC waverider.}
    \label{fig:midplanehighorder}
\end{figure}
Skewness and kurtosis shown in figure \ref{fig:midplanehighorder} further support the trends observed in figure \ref{fig:midplanetrends} and figure \ref{fig:midplanestats}. The skewness for pressure is positive for the $X/L = 0$ and $x/L = 0.05$ locations and then becomes negative. The heat flux is negative for the $X/L = 0$ and $x/L = 0.05$ locations and then becomes positive at the other locations. Kurtosis has the highest negative value at $x/L = 0.05$ location, which further reinforces the pressure as the most sensitive to the variation of the Prandtl number at that location.

\subsection{Polynomial Chaos Expansion}
\autoref{tab:pressurepcecompare} shows a comparison of mean and standard deviation using two methods (1) direct calculation of mean and variance and (2) using the PCE algorithm described in Section II. The purpose is twofold, first to identify the sensitivity and second to validate our implementation of the PCE algorithm. In the current case, there is only one independent variable, $P_{tr}$ for which we can easily perform any statistical analysis manually. However, when the number of parameters increases, we will need a reliable algorithm to conduct this analysis; the excellent comparison between the two here validates our algorithm and will be used in subsequent investigations.

\begin{table}[h]
    \centering
    \begin{tabular}{c|c|c|c|c}
            
        x/L & Mean (PCE) & Mean (Direct statistics) & $\sigma$ (PCE) & $\sigma$ (Direct statistics) \\
        \hline
        0.0 & 672018.0  & 672019.1 & 40.02 & 43.05 \\
        0.05 & 26793.3 & 26793.3 & 12.3 & 13.2 \\
        0.1 & 22802.3 & 22802.1 & 24.7 & 26.5 \\
        0.2 & 20604.0 & 20603.7 & 41.9 & 45.0 \\
        0.5 & 20480.1 & 20478.7 & 165.3 & 177.9 \\ 
        0.9 & 23325.4 & 23323.1 & 428.2 & 460.7  
    \end{tabular}
    \caption{Comparison between mean and standard deviation of pressures at each point along top surface centerline, with 5th order PCE}
    \label{tab:pressurepcecompare}
\end{table}

See a comparison between 3rd and 5th order PCE analysis of the pressure at point x/L = 0.9 as shown in \autoref{tab:pressurepceordercompare},

\begin{table}[h]
    \centering
    \begin{tabular}{c|c|c}
            
        Order & Mean & $\sigma$  \\
        \hline
        3rd order & 23315.3 & 421.5 \\
        5th order & 23325.4 & 428.2 \\ 
    \end{tabular}
    \caption{Comparison between a 3rd and 5th order PCE analysis}
    \label{tab:pressurepceordercompare}
\end{table}



\section{Summary and Conclusions}

Turbulence parameters for RANS turbulence models are constants that are typically measured experimentally. As a result, they contain uncertainty in their estimations, with many hypersonic codes today using those specific constants as defaults, which may not properly address turbulence in hypersonic flow scenarios. In this work, direct statistical analysis is utilized to show how one such variable effects the outcome of the simulation, The turbulent prandtl number has a more significant contribution to the output pressures and heat flux depending on body measurement location. Polynomial chaos expansion techniques are employed as an example that it could be used to predict these complex systems. Future work will require not only a single uncertain input, but rather multiple inputs, which may not be independent from one another. PCE methods are extremely valueable to such problems as they can consider correlated, dependent variables. 


\section*{Acknowledgments}
This research was supported by DEVCOM Army Research Laboratory grants, W911NF-22-2-0058 and W15P7T-19-D-0126. Jeremy Redding is a PhD Fellow supported through the Army Educational Outreach Program (AEOP) program with ARL, cooperative agreement W9115R-15-2-0001. Luis Bravo was supported by the 6.1 basic research program in propulsion sciences. The authors gratefully acknowledge the High-Performance Computing Modernization Program (HPCMP) resources and support provided by the Department of Defense Supercomputing Resource Center (DSRC) as part of the 2022 Frontier Project, Large-Scale Integrated Simulations of Transient Aerothermodynamics in Gas Turbine Engines. The views and conclusions contained in this document are those of the authors and should not be interpreted as representing the official policies or positions, either expressed or implied, of the DEVCOM Army Research Laboratory or the U.S. Government. The U.S. Government is authorized to reproduce and distribute reprints for Government purposes notwithstanding any copyright notation herein.

\bibliography{refs/sample.bib}

\end{document}